# Dynamic and rate-dependent yielding behavior of $Co_{0.9}Ni_{0.1}$ nanocluster based magnetorheological fluids

Injamamul Arief[1], Rasmita Sahoo[2], P.K. Mukhopadhyay[1(a)]

[1]LCMP, Department of Condensed Matter Physics and Material Sciences, S. N. Bose National Centre for Basic Sciences, Salt Lake, Kolkata 700 098, India.

[2]School of Physics, University of Hyderabad, Hyderabad 500046, India.

**Abstract:**

In this paper we performed steady shear and oscillatory magnetorheological (MR) studies in magnetic fluids containing CoNi nanoclusters of 450 nm in diameter. Co-rich nanoclusters were synthesized by conventional homogeneous nucleation without any external surfactant or reducing agent in liquid polyol at elevated temperature. The x-ray diffraction, energy dispersive X-Ray analysis, scanning and transmission electron microscopy studies were done for analyzing the sample composition and morphology. Two variants of fluid samples were prepared by dispersing 15 vol% and 20 Vol% of CoNi powders in castor oil. Room temperature steady magnetoshear studies indicate viscoplastic behavior with stronger dependence of static yield stress on magnetization than a dipolar coupling that was operational in the dynamic yield stress. Magnetosweep measurements at constant shear rate showed interesting relaxation at high magnetic fields. We also explored dynamical elastic behavior through oscillatory magnetorheological studies under both strain sweep and frequency sweep modes, and showed glass transition like phenomenon occurring in them above critical shear amplitudes.

[(a)]Email: pkm@bose.res.in

## 1. Introduction

Magnetorheological (MR) fluid constitutes a smart material whose flow behavior can be tuned by application of external magnetic field. It consists of magnetic particles of micron to submicron (0.01-10 µm) dimensions, dispersed in carrier liquid [1-2]. The response of fluid under magnetic field is attributed to anisotropic, three-dimensional chain-like structures formed parallel to the direction of applied magnetic field. Therefore, MR fluid exhibits changes in viscosity by several orders of magnitude in presence of field [3-4]. As carrier liquid is entrapped in between the fibrous columnar aggregates of magnetized particles, the effective solid phase concentration increases, therefore, the on-state viscosity jumps up manifolds. The field-induced columnar structure can withstand external stress and has a potential for future magnetomechanical applications [5]. MR fluids have already been utilized in semi-active devices such as loudspeaker and microphone cone centering, tunable dampers and clutches [1-5].

Ferromagnetic particles are normally used in magnetic suspension for their higher magnetic saturation and permeability. Ferrofluidic systems are colloidal dispersions of superparamagnetic ultra-small (~10 nm) nanoparticles with minimum polydispersity and long term stability [3]. However, very low field-induced mechanical stress limits their usage [6]. An MR fluid, on the other hand, utilizes micron-sized ferromagnetic particles with moderate-to-low suspension stability and has high magnetomechanical stress. A lot of researches have been reported dedicating the improved MR effect, as well as enhanced fluid stability against settling [7-8]. In contrast to conventional carbonyl iron particles (CIP), ferromagnetic binary alloys provide a more viable alternative for their higher saturation magnetizations, low coercivity and remanence and excellent control of shapes and sizes. Previously, CoNi nanosphere and nanowire-based ferrofluids were reported by Lopez-Lopez et al.[9] Cubic and spherical FeCo-based MRFs and CoNi nanoflower-based MR fluids were also investigated by the authors [10-11]. The present work is aimed at investigating MR properties of CoNi-based MRFs under oscillatory and steady shear modes. Cobalt rich $Co_{0.9}Ni_{0.1}$ nanoparticles (~450 nm) were synthesized by standard polyol reduction method. Two different MR samples of different volume fractions were prepared by dispersing CoNi into castor oil. We investigated the mechanism that governed size distribution of field-induced internal chain structures.

## 2. Experimental details

The synthesis of spherical $Co_{0.9}Ni_{0.1}$ nanoclusters was carried out by typical one-pot polyol reduction method. In this process, precursor solution was prepared by dissolving nickel (II) acetate tetrahydrate (≥98%, Sigma-Aldrich) and cobalt (II) acetate tetrahydrate (≥98.0%, Sigma-Aldrich) in ethylene glycol (Merck) with 1:9 molar ratios. For the synthesis of $Co_{0.9}Ni_{0.1}$, total molar concentrations of the metals were fixed at 0.5M. Ethylene glycol (EG) served both as reducing agent and surface binding material. Afterwards, EG solution of sodium hydroxide (10 mL, 1.5 M), previously purged with argon, was added drop-wise to the precursor under constant stirring. The resulting solution turned bluish-green due to the formation of mixed metallic double hydroxide, i.e. homogeneous solid phase. The suspension was then refluxed under nitrogen blanket at $200^0$C for 2 hours. The purified $Co_{0.9}Ni_{0.1}$ suspension was separated by centrifugation and subsequent washing with Millipore® water, acetone and absolute ethanol, then the suspension was dried in air at $60^0$C to obtain the powder.

Particle morphology, size and shape were investigated by a field emission scanning electron microscope (Quanta FEG®, FEI) with an energy dispersive x-ray (EDAX) attachment, and transmission electron microscope (Technai F20®, FEI). Crystal structures and phases of the powdered samples at room temperature were identified powder x-ray diffraction by using a PANalytical X'Pert PRO® diffractometer (scan time was 18 min within $2\theta = 40^0$ to $85^0$) using monochromatic Cu-$K_\alpha$ radiation ($\lambda$=0.51418 nm). Room temperature magnetometric study of sample pellets (pressed powder) was performed on a Lakeshore Cryotonics® Inc. model 7400 VSM.

The MR fluids were made by dispersing proper amounts of this powder into castor oil (viscosity 0.879 Pa.s at $25^0$C) through mechanical stirring and ultrasonication. Two MR suspensions (MR1 and MR2) were prepared, with dispersed phase concentrations of 20 vol% and 15 vol%, respectively. Silica nanopowder (Merck, 2 vol% for each suspension) was also added to avoid irreversible particle aggregation and settling.

Both on- and off-field magnetorheological measurements with were investigated using a commercial rheometer (Anton Paar MCR Physica 501®) with magnetorheological attachment (MRD 170®) in strain-controlled mode and at room temperature. The parallel plate system with plate diameters of 20 mm was used for all measurements. A fixed plate-gap of 1 mm was maintained throughout the measurements. The magnetic field was generated vertically with respect to the direction of flow. For magnetorheological measurements, magnetic field was varied from 0 to 0.6 T. Before any measurement, the sample was pre-sheared at 20 $s^{-1}$ for about 30 s. In magnetosweep experiment, field was varied from 0 to 1.1 T and back to zero under constant shear rate of 10 $s^{-1}$. Dynamic (oscillatory) magnetorheological properties were probed by amplitude and frequency sweep experiments at $25^0$C. In amplitude sweep, an oscillatory strain ranging from 0.01% to 100% was applied to the sample under constant frequency of 10 Hz. The measurements were repeated at 0.04 T, 0.24 T and 0.5 T. Frequency sweep measurements were carried out under constant strain amplitude of 0.02% under the fields of 0.04 T, 0.24 T and 0.5 T. Angular frequency sweep was performed from highest (100 Hz) to lowest (0.1 Hz) frequency.

## 3. Results and discussions

### *3.1 Structure and morphology of magnetic nanoclusters*

The crystal structures and phases of the alloy nanoclusters were determined by powder x-ray diffraction as shown in Fig. 1. The featured peaks at $\theta$= $44.6^0$, $51.7^0$, $77.6^0$ were assigned to fcc crystal structure while significant presence of hcp Co phases at $\theta$ =$42^0$, $44.8^0$ and $47.5^0$ were also reported. It was previously observed that Co-rich CoNi alloys tend to crystallize in mixed fcc and hcp phases [11-12]. However, all peaks correspond to fcc phases were indexed to fcc Ni (JCPDS 15-0806) and Co (JCPDS 01-1260). According to Fiévet et al. [13], CoNi alloyed nanostructures with more than 80 atomic% of Co exhibited higher probability of stacking fault. The surface of nanoclusters appeared to be rough for Co-rich CoNi due to extended growth step by aggregation of primary nanoparticles units.

The morphologies and average sizes of the nanoclusters and size distribution were obtained from FESEM and low magnification TEM images and are shown in Fig. 2. Average size was calculated to be 450 nm. The surface morphology of $Co_{0.9}Ni_{0.1}$ can tentatively be related to the mechanism of formation. It follows two distinct phases of nanocrystal formation: nucleation and growth. The nanoparticles in polyol can either follow stepwise addition of metal atoms to form seed and subsequent coalescence of seed into nanoparticles, or primary nanoparticles aggregate to form larger nanoclusters. It was seen that in this case, the latter arrangement takes place [11-14]. The selected area diffraction (SAED) pattern for the material indicates polycrystalline nature of sample (Fig. 2D). The EDX spectrum of the as-prepared $Co_{0.9}Ni_{0.1}$ nanocluster is shown in Fig. 2 (C). It shows that Co/Ni ratio in the nanocluster is very close to the initial molar ratio (1:9 for [Ni]: [Co]) of the metals salts. The spot EDX analysis for different position across the diameter of nanocluster was also performed and was observed that $Co_{0.9}Ni_{0.1}$ obtained compositional homogeneity.

Magnetic hysteresis at room temperature and FC-ZFC magnetization curves of the nanoclusters are shown in Fig. 3. The M vs. H hysteresis loop indicates ferromagnetic character of the sample at room temperature. The saturation magnetization ($M_s$), remnant magnetization ($M_r$) and coercive field ($H_c$) values were calculated to be 119.5 emu/g, 21 emu/g and 142 Oe, respectively. Magnetic saturation for $Co_{0.9}Ni_{0.1}$ was lower than that of bulk Co or $Co_{80}Ni_{20}$ alloy [12]. As reported earlier, $M_s$ values for nanoparticles are considerably smaller than that of bulk alloys with same compositions, possibly due to the presence of passive dead layer of metal oxides [12-13]. The ratio of $M_r$ to $M_s$ at room temperature was 0.17 which is considerably lower than 0.35, indicating randomly oriented, blocking type nanoparticles [15]. To confirm the magnetic state, ZFC and FC measurements from 75K to 350K, were performed in presence of field (100 Oe). The bifurcation between these indicates irreversibility starting from room temperature, due to the magnetic nano particles.

### *3.2. Magnetorheological characterizations*

#### *3.2.1. Steady shear magnetorheological properties*

The flow curves (Fig. 4) under controlled shear rate modes were obtained as a function of different magnetic fields for both magnetorheological samples, MR1 and MR2. The magneto-rheograms for both MR1 and MR2 characteristically displayed typical MR behavior. It is evident that a threshold shear stress is required to make the initial flow, which is a property of yield stress fluid and this initial stress at very low shear rate (~0.001-0.1) is termed as static yield stress ($\tau_{ys}$) [2,6,9]. With increasing field, initial stress required to make the fluid flow would also go higher, owing to the formation of more robust field-dependent interparticle network. However, values of shear stresses obtained in different magnetic fields were different for both MR1 and MR2. Increasing concentration of suspended magnetic particles usually enhances MR performance [16].

Generally, field-dependent magnetorheological parameters (yield stress, shear viscosity, storage modulus) exhibit linear or power-law relationship with concentration of MR fluids [17]. This also holds good for the samples as we observe systematic increment of shear stress upon application of field. In the beginning, the field induced change was high, but at higher fields (0.4 T or more), the changes leveled off as magnetic particles tend to saturate progressively. With increasing field, MR1 with higher magnetic concentration tended to form plateau-like rheograms, giving rise to more shear-thinning behavior beyond a critical shear stress. Transition from solid-like to liquid-like state was more regular for MR2 with lesser magnetic vol%. For concentrated fluids, due to flocculated structures formed at rest, flow of fluid was hindered below a certain shear stress. Beyond that stress (yield stress), it started flowing suddenly. For the lower concentrated fluids, transition from solid to liquid state was more gradual as structure at rest was relatively weaker. However, another way to look at this behavior is to define the flow characteristics of magnetic fluid through a parameter $\eta_M$, called magnetoviscous effect (MVE). This is defined as the change in shear viscosity at a particular field and shear rate, in comparison to its zero field viscosity [9]:

$$\eta_M = \frac{\eta_H - \eta_0}{\eta_0}$$

Where $\eta_H$ and $\eta_0$ are the viscosities of MR fluid at a specified magnetic field and zero magnetic field, respectively. Fig. 5 illustrates the MVE in flow regime as a function of shear rate for MR1 and MR2 under different magnetic fields. The magnitude of MVE decreases systematically with increasing shear rate for both samples and at all fields. This is generally expected as in flow regime hydrodynamic interactions must overcome magnetostatic forces. Therefore, field-induced aggregates are broken to continue the flow. It is also observed that MVE is more pronounced in MR1, compared to MR2. The relative increment of viscosities at a given field and a given shear rate for MR1 is higher than that of MR2 due to higher off-state viscosity of the former. Moreover, higher concentration of magnetic nanoclusters in MR1 facilitated stronger field-induced chain structure, making it more rigid to external shear.

*3.2.2. Magnetosweep studies and field-dependent relaxation*

Fig. 6 shows magneto-sweep curves for the two samples. At a constant shear rate of 10 $s^{-1}$, viscosity was varied as a function of magnetic field within a range of 0 to 1.15 T. It was observed that viscosity increased with increasing field, owing to the formation of field-dependent microstructures. At higher field, MR suspensions tended to saturate and viscosity remained unchanged for the rest of the field. Although both samples contained same nanocluster with $Co_{0.9}Ni_{0.1}$ compositions, saturating field limit for MR1 was higher than that of MR2. At any field, viscosity for MR1 was found to be much higher than that of MR2, as is usually observed in suspensions with higher particle volume fractions. The tendency to form lateral aggregation (known as zippering) between single chain microstructures was more pronounced in MR1. For both samples, viscosity corresponding to backward field sweep was higher than forward sweep. This gave rise to hysteresis behavior [3]. This is a behavior that is similar to M-H curves, and indeed on comparison to Fig. 3, they look alike. This means that the dipole-dipole interaction was the main cause of field induced viscosity here, and once the systems were saturated, they did not offer further hindrance to the flow. The major change was in the final value of viscosity, within the magnetic fields, these are 3 times more for the denser sample. Since the volume fraction is about 33% more, this increase may be expected.

The interesting observation is the crossovers of the field reverse curves below the field forward curves. Since the carrier fluid was non magnetic, the response of the fluid must be entirely due to magnetic particles. So the hysteresis was not expected to show any new feature. We found one report of such measurements (i.e. viscosity during both increase and decrease of field) in the literature.[18] However, apart from obvious difference in behavior due to change in chemical nature of sample, they did not observe this cross over too. However, they did not impose saturation field like us, and we think – in the absence of any plausible explanation, this holds the key to the difference in behavior.

Looking minutely in the viscosities at final high magnetic fields, it is seen that there was a waiting time before the field direction was reversed. The fall in viscosity happened during then. While it is tempting to think that it was the fatigue, we think this was more likely due to rising local temperature effect. Since the system spent some more time there at the same place, while under constant shear, and the viscosity was very high, so the liquid had some (local) temperature rise. Even if that must be a tiny effect, we found from the zoomed region (inset of Fig. 6) that there indeed was a fall in both cases at the extreme right field, thus corroborating our view. Once the field was reduced, the viscosity was restored to the value corresponding to 25°C, but this took some time. The relaxation went on for some time till the values matched the rising values, and then the memory effect took over. The structure was stronger in the return leg of the journey, and so the viscosity was higher.

*3.2.3. Static and dynamic yield stress:*

Static yield stress ($\tau_{ys}$) is defined as the minimum non-zero shear stress required to break the field-induced particle chain structures formed under influence of magnetic field and initiate the flow. Therefore, it is an important magnetorheological parameter often used to characterize MR fluid's strength. While theoretically it is the value at zero stress rate, it is not practicable to measure it this way. Rather, it is taken from the value of shear stress at lowest measured shear rate ($10^{-3}$-$10^{-1}$ $s^{-1}$). On the other hand, dynamic yield stress ($\tau_{yd}$) is obtained by fitting the rheograms at higher stress rates with Bingham equation, $\tau = \tau_b + \eta_p \dot{\gamma}$, where $\eta_p$ is plastic viscosity [9]. Therefore, this is the extrapolated value at which the plastic flow would have occurred first. Since this is the major portion in the strain-strain rate curve in which the intended device will operate, it is also a very important parameter. The representative graph of static and dynamic yield stresses for MR1 and MR2 as a function of magnetic fields is shown in Fig. 7. As expected, $\tau_{yd}$ are always higher than $\tau_{ys}$, and within the measurement range, they increased monotonically with applied field.

To understand the origin of both types of yield stresses in the MR fluid, we tried to find the relationship with magnetic field. Ginder's law is such a well-known relation that predicts a power-law behavior for yield stress versus magnetic field (H) [19]. This is plotted in Fig. 8. In this case we find that an $H^{3/2}$ dependence of yield stress was obtained up to an intermediate field region but that broke down in higher field. This was observed before [10]. There is an alternative relation to the fit. According to Klingenberg et al. [20], a better fit can be observed when H is replaced by M and it is applicable to wide range of M. This is plotted in Fig. 9, with both types of $\tau$ and M values calculated from powder magnetization data (M-H curves, vide Fig. 3). It can be seen that while the $\tau_{yd}$ values fitted reasonably with $M^2$ for both the concentrations, the static values matched better with $M^3$, a higher power than with $M^2$. This behavior can be explained as follows. The fit to $M^2$ can be considered due to dipole–dipole interactions among the particles. This was the basic assumption in that paper [20]. However, for the static case the interaction was stronger. This happened because at the lowest shear rate, the particles were predominantly arranged in columnar fashion, and therefore multipolar contribution in the interaction was expected. In the other case, the fluid had already reached plastic limit and the columnar structure was disrupted by then. Therefore interaction among the individual magnetic particles became mostly dipolar type.

*3.2.4. Oscillatory magnetorheology:*

Oscillatory dynamic measurements are performed to explore the dynamic viscoelastic behavior of the fluids. Both amplitude and frequency sweeps were carried on the two samples under the same three magnetic field values. In case of amplitude sweeps, samples were subjected to constant frequency (10 Hz) sinusoidal deformations and responses were recorded various oscillatory shear regions over a period of time. Linear viscoelastic regime (LVR) was observed at a very small strain amplitude ($\gamma_0 \approx 10^{-1}$-$10^{-3}$) where the viscoelastic moduli were largely independent of strain amplitudes. However, as stress amplitude was increased beyond a critical strain ($\gamma_c$), these moduli became dependent on strain and drooped [21]. The behaviors are shown in Fig. 10 (A) and Fig. 10 (B).

The dependence of the moduli on magnetic field is obvious. $G'$ and $G''$ increased as the field was increased, however, initially there was an order of magnitude increase in the moduli when the field was varied by 6 fold, but the difference steadily became less as the strain was increased, implying loss of elasticity. One interesting difference between the behavior of the two liquids is that in case of MR1 (sample with higher concentration), the linear viscoelastic range is not very clear, unlike in case of MR2. This apparent ambiguity can be explained by the fact that the structures formed under higher fields were more prone to be affected by strain amplitude. The plateau at linear regions appeared not because of systematic deformation of particle chain junction, but rather microscopic movements and minor twisting within the chain network. It is interesting to note that at high strain, second pseudo-plateau regions of $G''$ appear for both fluids. This quasi-linear viscoelastic region signifies the homogeneous rupture of particle networks and an eventual increment in loss modulus ($G''$). For MR1, second pseudo-plateau appeared to be less prominent with increasing field. Beyond this, a crossover point ($G' = G''$) appeared marking transformation into flow regime. This is akin to glass transition observed for polymer melts and the like. This was where the systems went over from gel like to liquid like structure. A further important observation is the bunching of $G'$ and $G''$ in this liquid region. This was only natural, for now the flow behavior is liquid like.

Frequency-dependence of the viscoelastic moduli for MR1 and MR2 was performed at a fixed strain, $\gamma$ of 0.02% (Fig. 11). This was much below the range of critical strain, $\gamma_c$. As found in the previous experiment the range was 0.05 to 0.5%. These measurements provide information about the effect of interactions among particles in the fluid. In linear viscoelastic regime (small deformation, $\gamma \ll 1$), microstructures retain strongly elastic behavior contributed by percolated and free suspension aggregates. At higher particle concentration (MR1) and field, structural rigidity of percolated aggregates can be explained by cage formation [22], i.e. particles formed rigid cage-like structures which accounts for frequency-independent G′. The contribution of matrix fluid was also responsible for linear dependence of G″ at higher driving frequency [23]. Coupling these information, we conclude that the system showed behavior expected from a structured or solid-like material.

## 4. Conclusions

In this paper, we have presented detailed magnetorheological characterizations of two MR samples at 25$^0$C with different $Co_{0.9}Ni_{0.1}$ nanocluster concentrations of 20 and 15 vol%. It was observed that MR fluid with higher particle concentration produced enhanced magnetorheological effects and higher field-dependent yield stresses. Steady shear magnetorheological studies were done to determine static and dynamic yield stresses under different magnetic fields. The behavior of these yield stresses were found to depend differently on average particle magnetization, with the static yield stress dependence on M being stronger of the two. This was interpreted as an indication of multipolar behavior. Magnetosweep studies showed up an interesting relaxation phenomenon that was interpreted as an effect of local heating. Finally oscillatory dynamic elastic studies revealed glass transition in the systems with high enough strain.

## Acknowledgments

One of the authors, IA thanks CSIR, India for the award of Senior Research Fellowship (SRF). The authors also thank Dr. Surajit Dhara, Associate Professor, School of Physics, Central University of Hyderabad, Andhra Pradesh, India for helping with magnetorheological measurements.

## Figures with captions

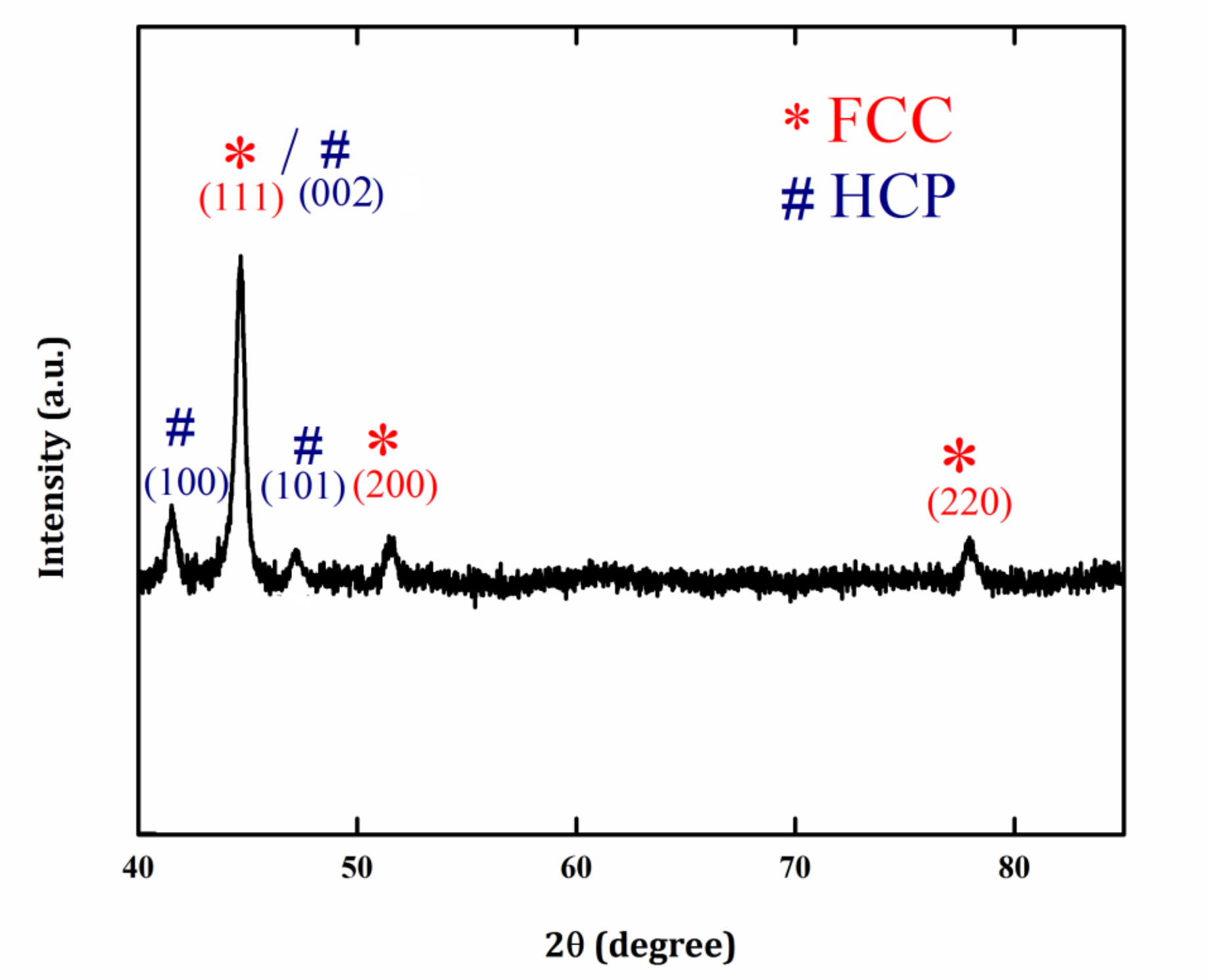

**Fig. 1.** Powder X-ray diffraction peaks of as-synthesized $Co_{0.9}Ni_{0.1}$ nanoclusters. All peaks were assigned to fcc and hcp phases.

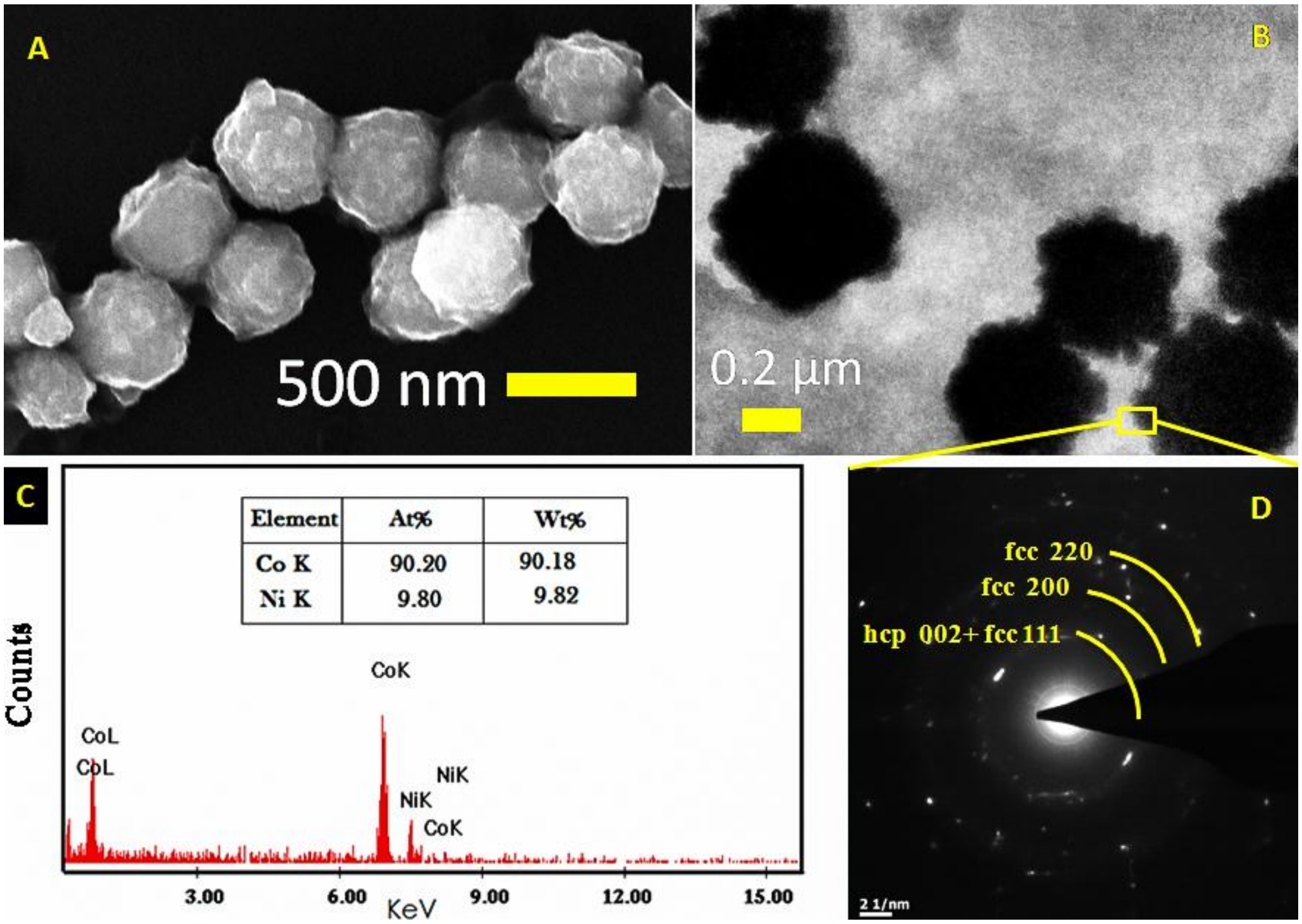

**Fig. 2.** (A) FESEM image of $Co_{0.9}Ni_{0.1}$ nanoclusters, average diameter of nearly spherical nanoclusters was found to be 450 nm; (B, D) low resolution TEM image and selected area diffraction pattern (SAED), (C) EDX of the sample confirmed the composition.

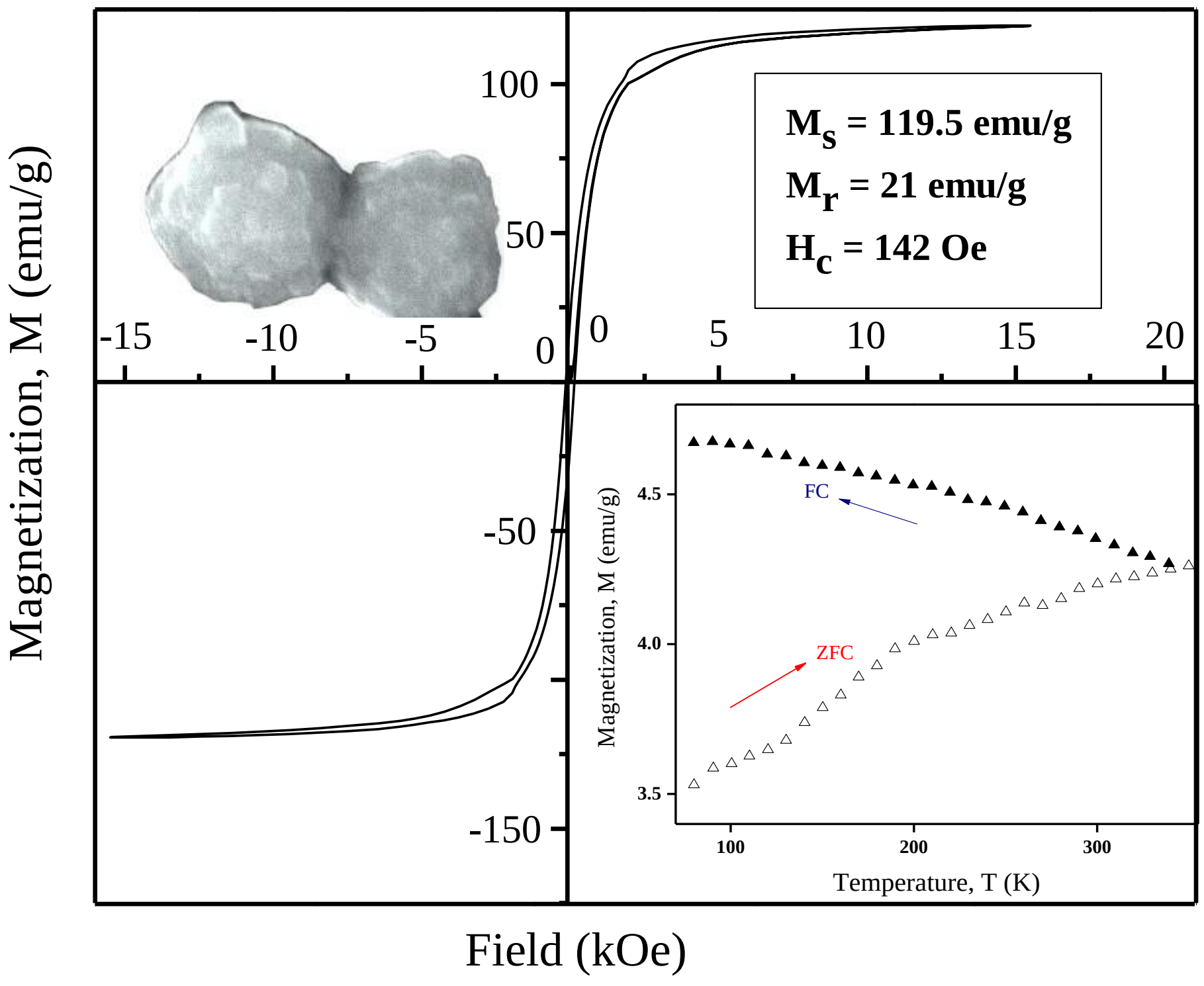

**Fig. 3.** Magnetization curve at T = 298K for $Co_{0.9}Ni_{0.1}$ nanoclusters. In the right inset, ZFC-FC magnetization curves under an applied field of 100 Oe were shown.

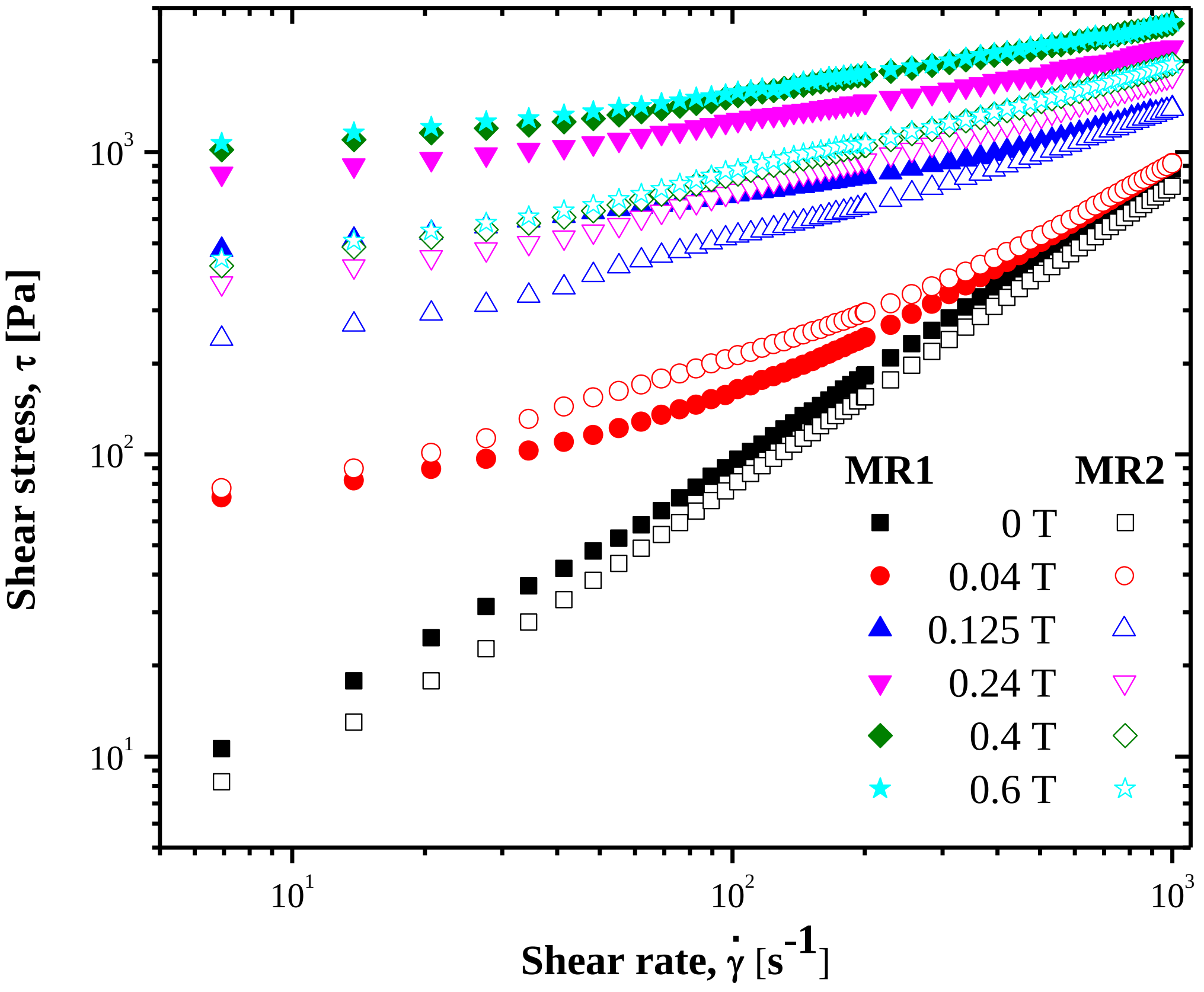

**Fig. 4.** Shear stress (τ) plotted as a function of shear rate ($\dot{\gamma}$) for different values of magnetic flux. Closed symbols is for sample MR1 whereas open symbol represents MR2.

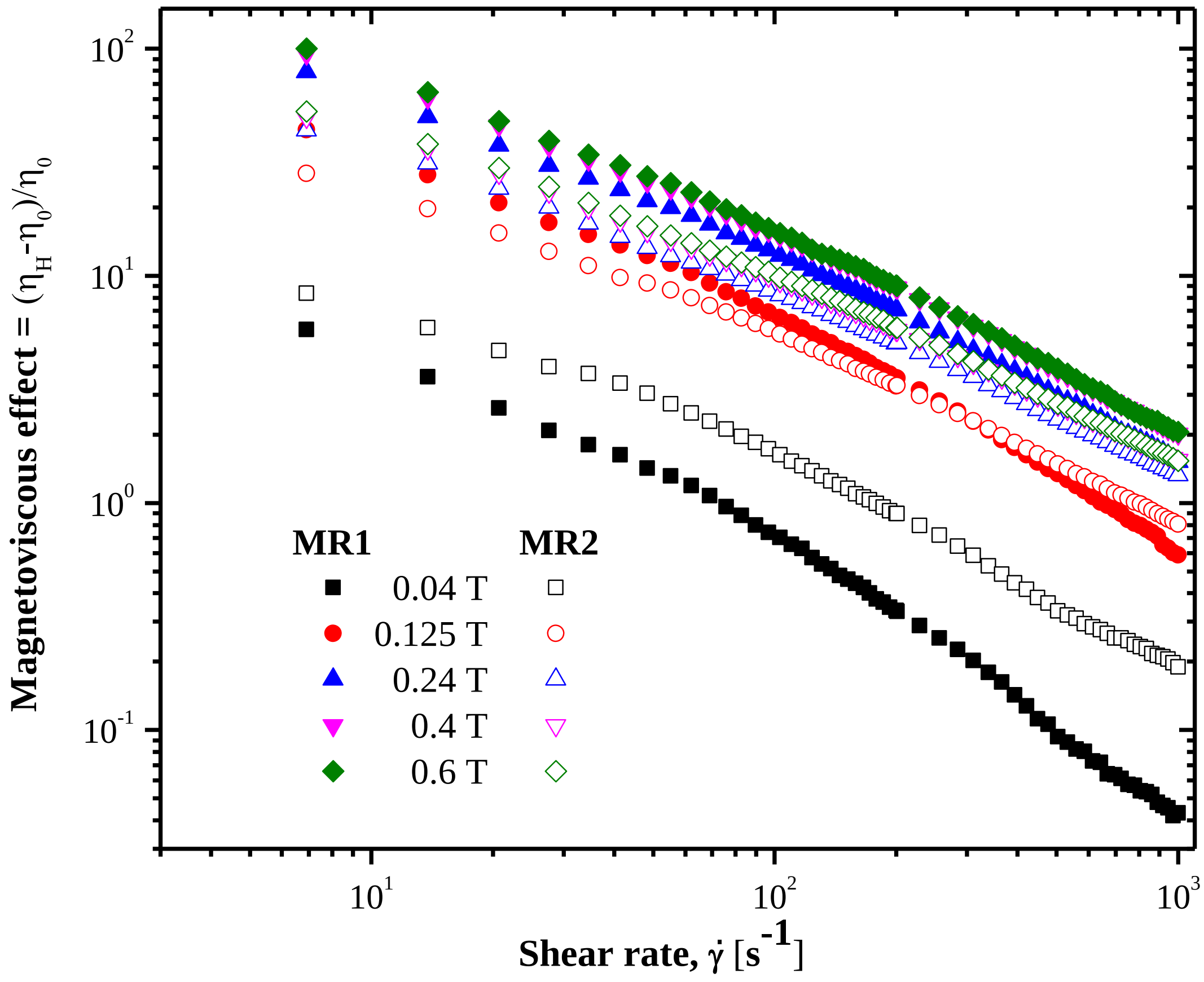

**Fig. 5.** Magnetoviscous effect as a function of shear rate for MR1 (solid symbol) and MR2 (open symbol) under different magnetic fields.

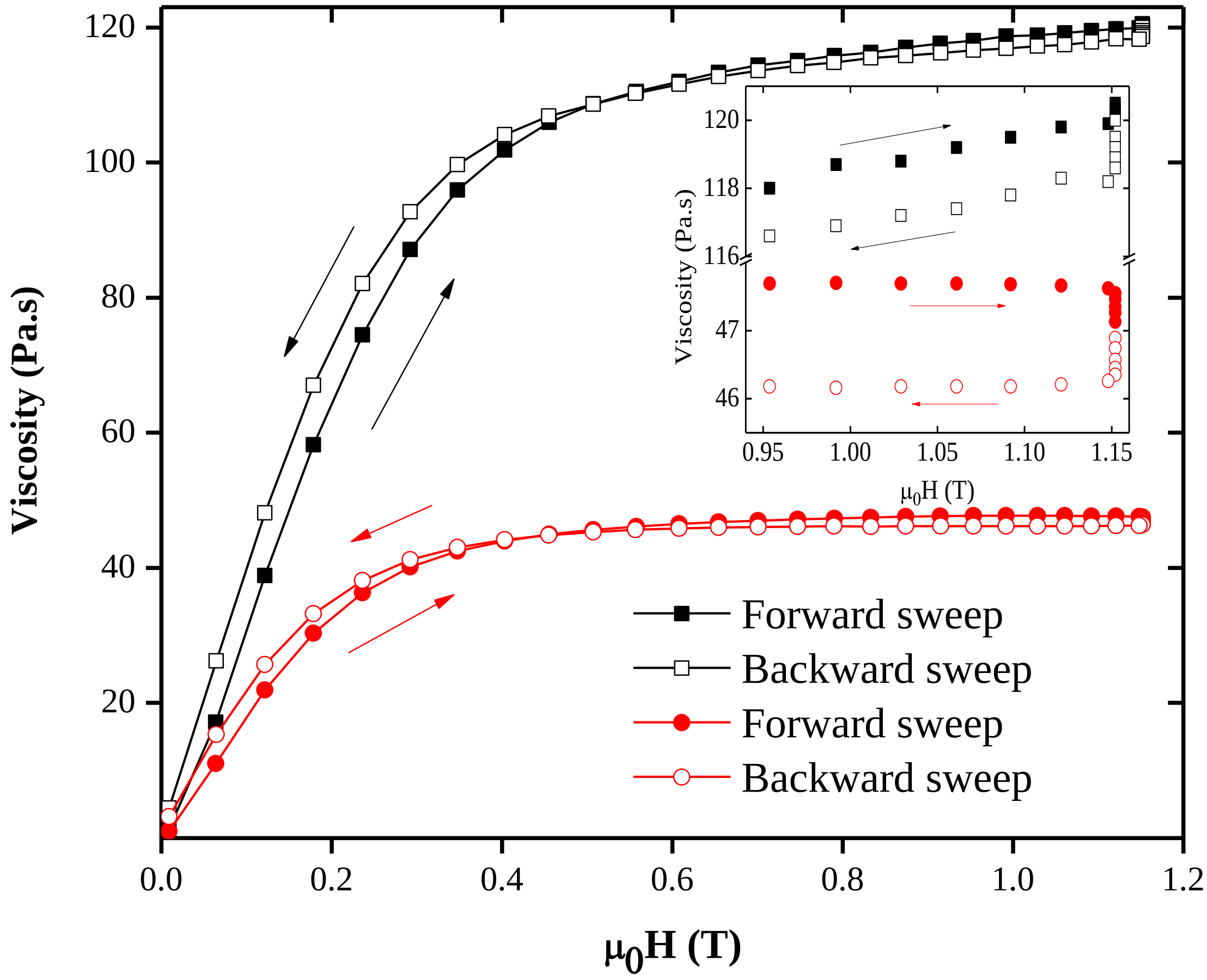

**Fig. 6.** Magnetic field sweep (viscosity vs. field) for MR1 (black square) and MR2 (red circle) at constant shear rate of 10 $s^{-1}$. Solid symbol represents increasing magnetic field, whereas open symbol stands for decreasing field. Inset, zoomed in portion of magnetosweep curves at higher fields.

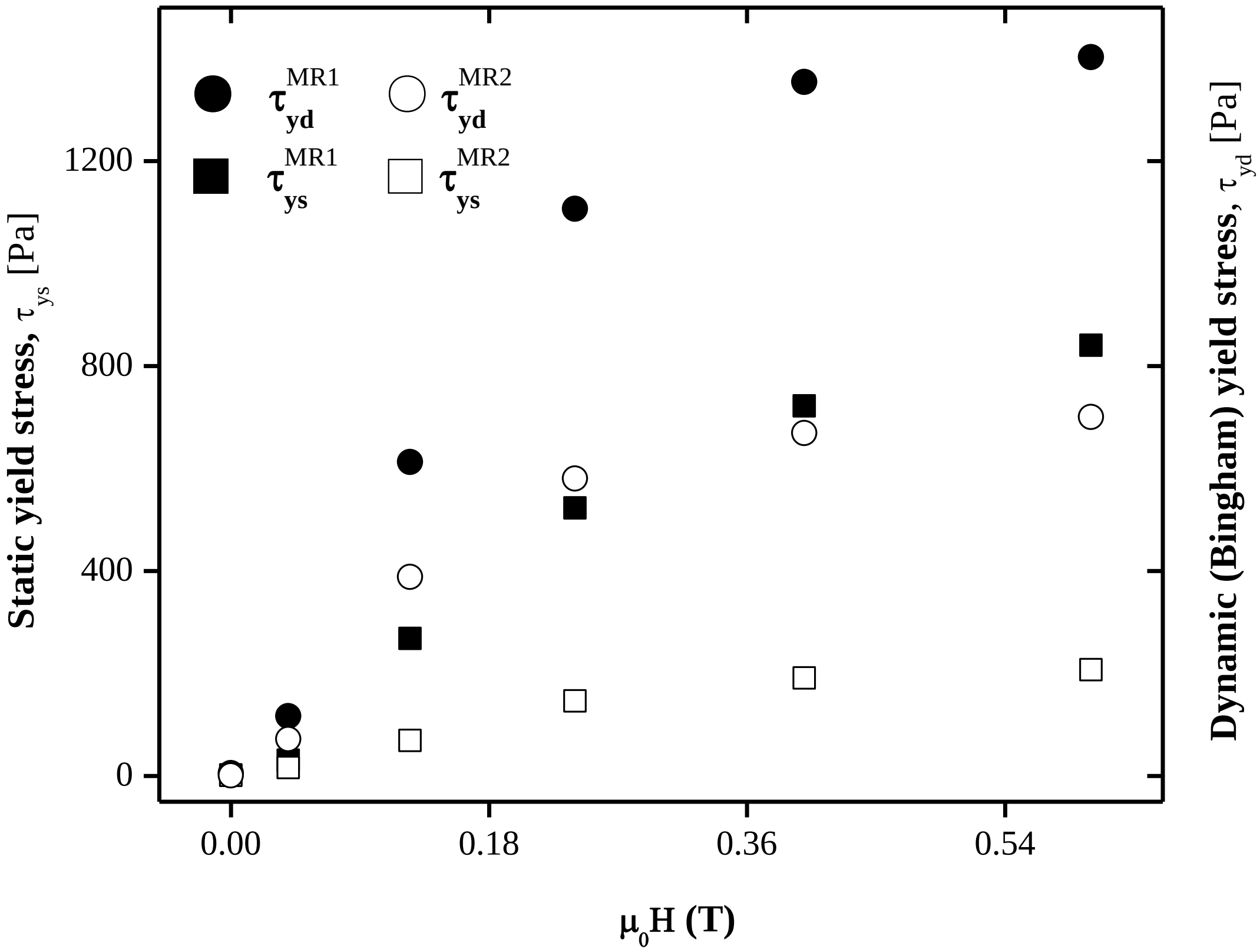

**Fig. 7**. Static ($\tau_{ys}$) and dynamic yield stress ($\tau_{yd}$) for both the samples MR1 (closed symbols) and MR2 (open symbols) are plotted as a function of magnetic field. Square symbols represent static yield stress whereas circular symbols denote dynamic yield stress.

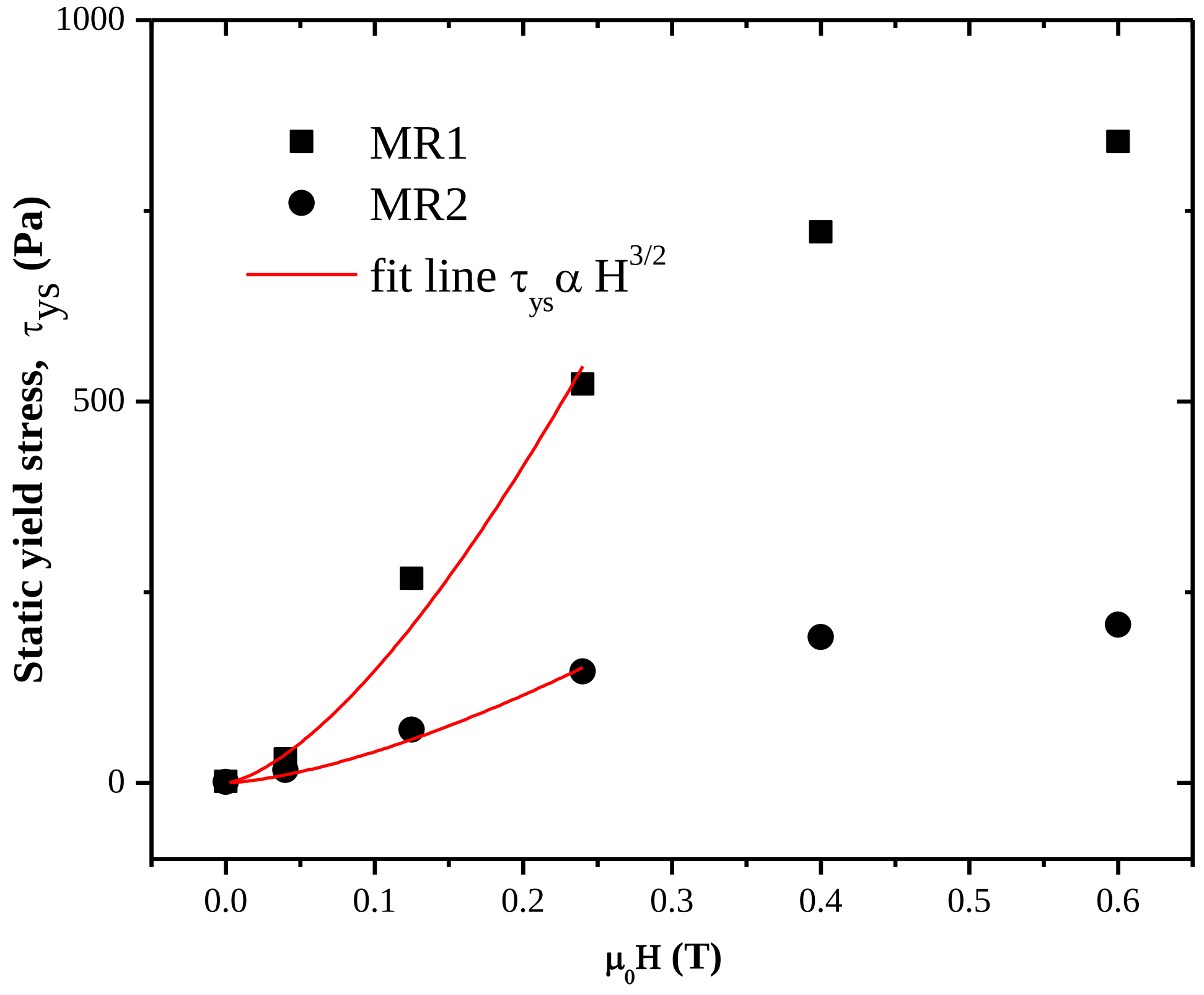

**Fig. 8.** Static yield stress versus magnetic field curves for MR1 and MR2. The fit lines represent Ginder's relation: $\tau_{ys} \propto H^{3/2}$

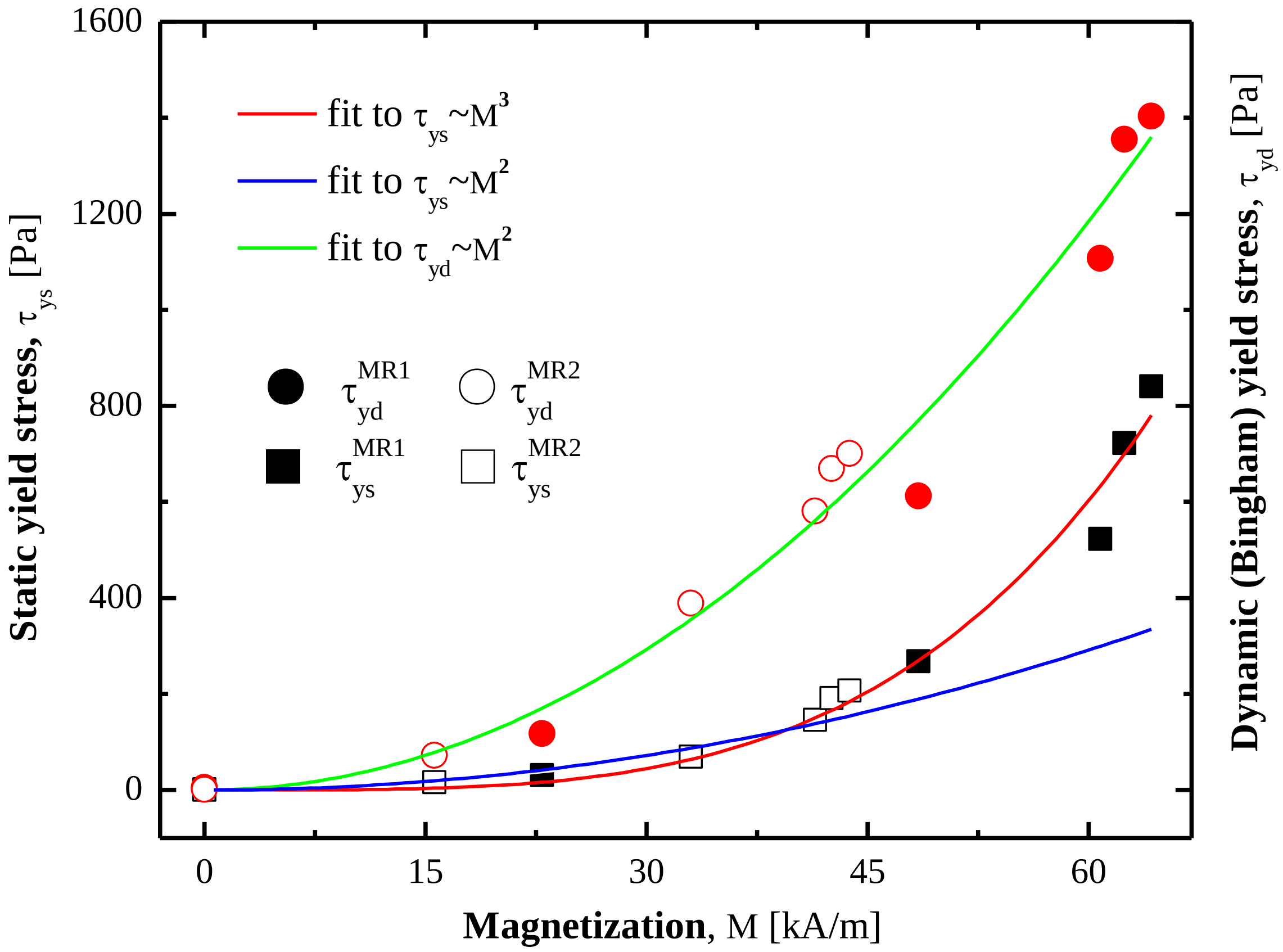

**Fig. 9.** Field-induced static (square) and dynamic (circle) yield stress for MR samples MR1 (solid symbol) and MR2 (open symbol) as a function of magnetization M.

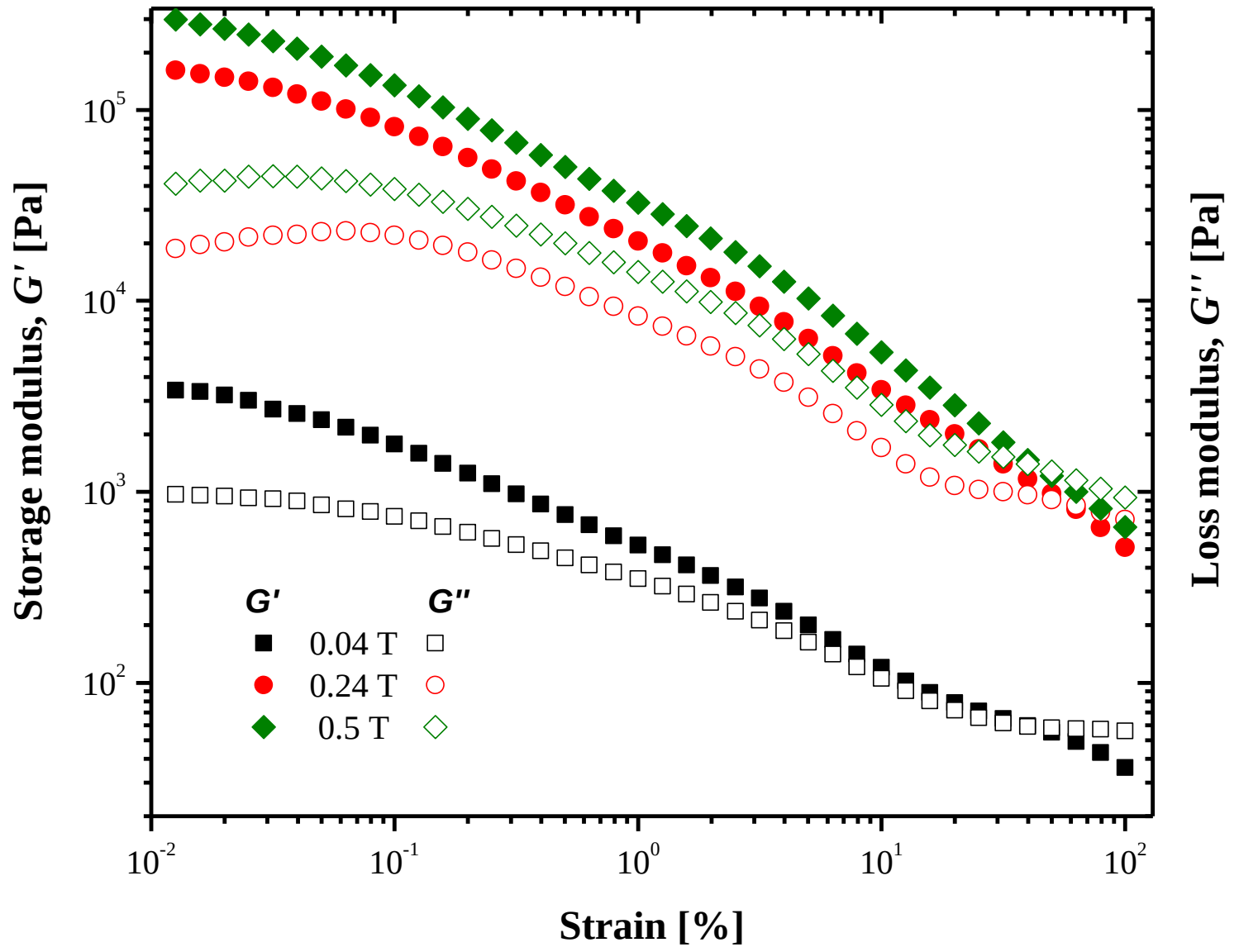

**Fig. 10 (A)**. Amplitude sweep of MR1 at a constant angular frequency of 10 Hz under different magnetic fields. Filled symbol indicates storage modulus (G′) whereas, open symbol is for loss modulus (G″).

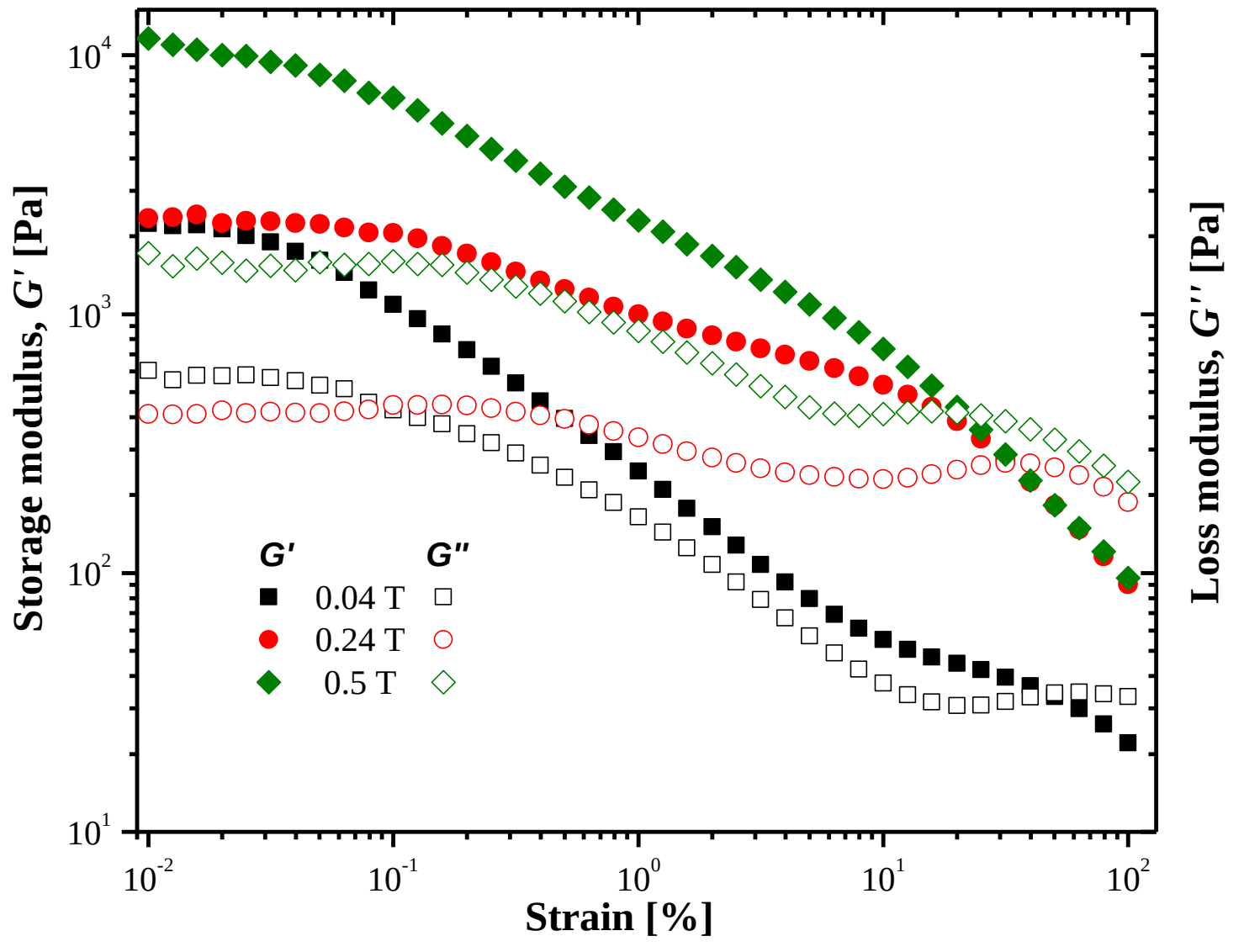

**Fig. 10 (B).** Amplitude sweep of MR2 at a constant angular frequency of 10 Hz and under different magnetic fields. Filled symbol indicates storage modulus (G′) whereas, open symbol is for loss modulus (G″).

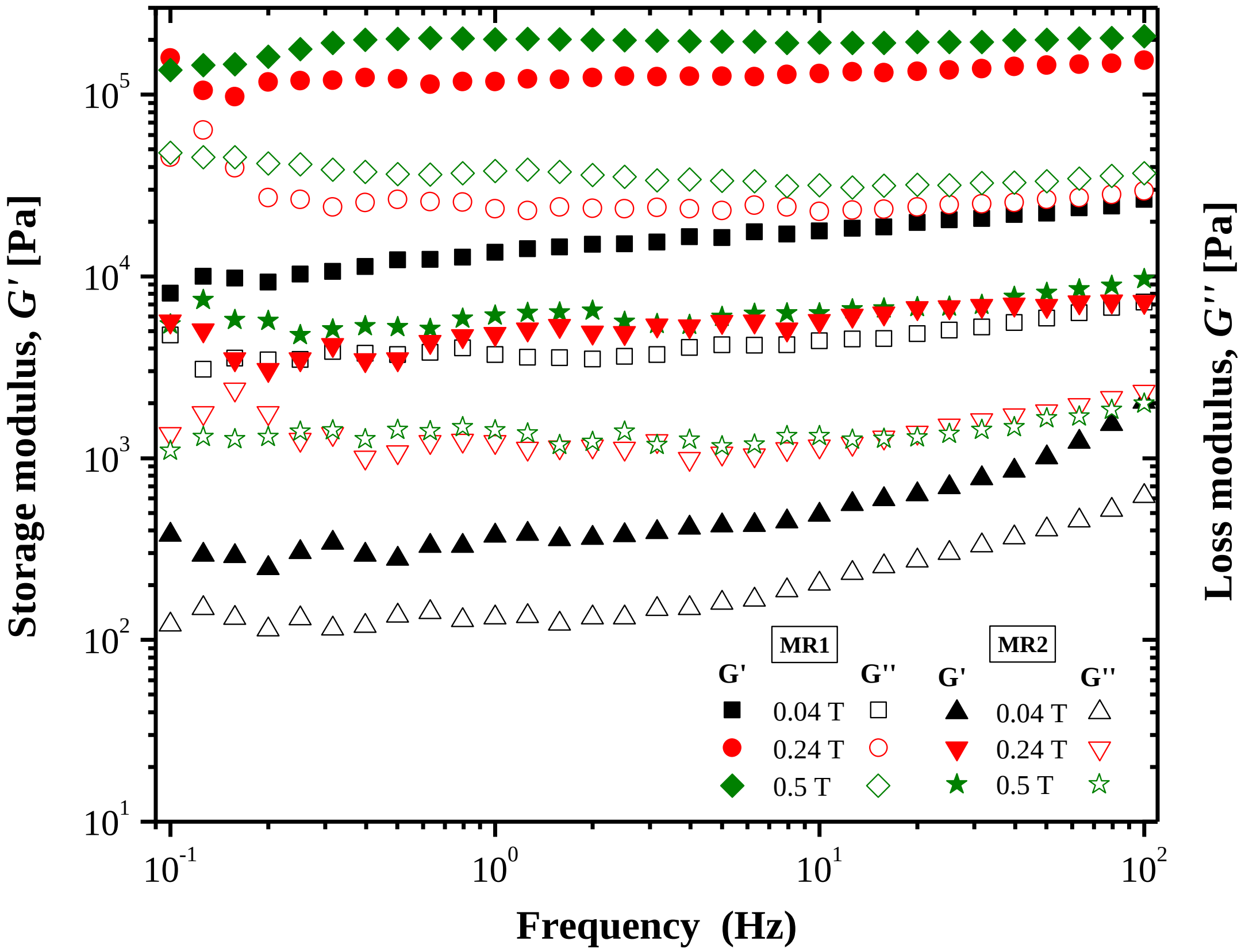

**Fig. 11.** Frequency sweep test at a constant strain amplitude of 0.02% for the samples MR1 and MR2 under different magnetic field. Solid symbol indicates storage modulus (G′) whereas, open symbol is for loss modulus (G′′).